\documentclass{article}

\usepackage{PRIMEarxiv}

\usepackage[utf8]{inputenc} 
\usepackage[T1]{fontenc}    
\usepackage{hyperref}       
\usepackage{url}            
\usepackage{booktabs}       
\usepackage{amsfonts}       
\usepackage{nicefrac}       
\usepackage{microtype}      
\usepackage{lipsum}
\usepackage{fancyhdr}       
\usepackage{graphicx}       
\graphicspath{{media/}}     
\usepackage{graphicx}
\usepackage[colorinlistoftodos,prependcaption]{todonotes}
\usepackage{regexpatch}
\usepackage{subcaption}
\usepackage{float}
\usepackage{mwe}
\usepackage{amsmath}
\title{Towards A Comprehensive Assessment of AI's Environmental Impact
}

\author{
  Srija Chakraborty \\
  USRA \\
  Washington, DC\\
  \texttt{\{schakraborty\}@usra.edu} \\
}

\begin{document}
\maketitle

\begin{abstract}
Artificial Intelligence, machine learning (AI/ML) has allowed exploring solutions for a variety of environmental and climate questions ranging from natural disasters, greenhouse gas emission, monitoring biodiversity, agriculture, to weather and climate modeling, enabling progress towards climate change mitigation. However, the intersection of AI/ML and environment is not always positive. The recent surge of interest in ML, made possible by processing very large volumes of data, fueled by access to massive compute power, has sparked a trend towards large-scale adoption of AI/ML. This interest places tremendous pressure on natural resources, that are often overlooked and under-reported. There is a need for a framework that monitors the environmental impact and degradation from AI/ML throughout its lifecycle for informing policymakers, stakeholders to adequately implement standards and policies and track the policy outcome over time. For these policies to be effective, AI's environmental impact needs to be monitored in a spatially-disaggregated, timely manner across the globe at the key activity sites. This study proposes a methodology to track environmental variables relating to the multifaceted impact of AI around datacenters using openly available energy data and globally acquired satellite observations. We present a case study around Northern Virginia, United States that hosts a growing number of datacenters and observe changes in multiple satellite-based environmental metrics. We then discuss the steps to expand this methodology for comprehensive assessment of AI's environmental impact across the planet. We also identify data gaps and formulate recommendations for improving the understanding and monitoring AI-induced changes to the environment and climate. 
\end{abstract}


\section{Introduction and Background}
Adopting artificial intelligence (AI), machine learning (ML) into Earth Sciences has enabled progress and accelerated scientific analyses, spanning a large range of research areas such as climate modeling, weather prediction, disaster monitoring and recovery, estimating greenhouse gas emission, land cover land use mapping, conservation, agriculture and food security, water quality analysis, to name a few. Use of AI/ML has derived insights into different Earth science processes of interest and accelerated scientific workflow by leveraging growing volume of data that is often heterogeneous such as remote sensing satellite data, climate models, digital twins, and in situ data. Moreover, Earth Sciences consist of complex datasets that exhibit high spatio-temporal variation, are often high dimensional, and multimodal. 
This necessitates the use of AI/ML to model large, complex datasets and monitor signals and processes of interest. Recently, foundation models have also shown promise in Earth Sciences in training on these large continually-growing datasets and fine-tuning to a range of downstream tasks~\cite{jakubik2023foundation, hong2024spectralgpt,noman2024rethinking}. As the Earth sciences are impacted by the lack of ground truth and labeled data~\cite{persello2022deep}, the use of foundation models allow generalizing to a variety of downstream tasks where gathering labels is more feasible. 

The rapid adoption of AI/ML in recent years is not characteristic to Earth sciences alone. Particularly, foundation models have resulted in a paradigm shift in AI due to its suitability to be adapted to a wide range of downstream use cases~\cite{devlin2018bert,brown2020language, bommasani2021opportunities}. Foundation model development is driven primarily by using massive volumes of data and leveraging enormous computational resources. Once, developed, these models can be fine-tuned for various downstream tasks, by using substantially lower amount of resources in a lightweight manner, for e.g. by using APIs. Thus, at present foundation model lifecycle begins with a small set of companies developing the most prominent and widely used models, followed by a downstream market of thousands of end user companies using lightweight inference mode versions~\cite{bommasani2023foundation}. 

Given the pervasiveness of AI/ML, there is a critical need for transparency around these models at various stages of the AI lifecycle for ensuring responsible use. 
Along with massive dataset sizes, the availability of massive volumes of compute through data centers and cloud have allowed the training of prominent models and is a key driver in enabling their performance~\cite{thompson2020computational, devlin2018bert, radford2019language, strubell2019energy,luccioni2023power}. This demand for compute resources puts a strain on natural resources and climate results in AI/ML's environmental impact, one which is not well quantified and monitored~\cite{crawford2021atlas, kaack2022aligning, crawford2024world}. The impact is likely to be higher around the training process. This is mostly done by a few prominent AI development companies whose collective environmental footprint is expected to be enormous, but is not transparently reported. Alternatively, while the environmental footprint is expected to be much lower at the inference stage, given the number of commercial applications built around prominent AI models running on publicly available cloud infrastructure, the impact at this stage is unlikely to be insignificant~\cite{dodge2022measuring, luccioni2023estimating, luccioni2023power}.

AI's environmental impact is multifaceted and includes its energy consumption and carbon footprint, water usage, and demand for raw materials. One of the relatively higher researched areas of this footprint is AI's electricity consumption. At present, several studies have highlighted AI's energy usage that is at the heart of training algorithms on massive volumes of data ~\cite{strubell2019energy, patterson2021carbon} and deriving inferences once trained~\cite{luccioni2023estimating, luccioni2023power}. These studies have focused on estimating electricity usage of specific natural language processing models~\cite{strubell2019energy}, and generative AI models~\cite{luccioni2023power}. Retrospective estimate of energy consumption may be challenging due to several run time specific factors that affect consumption such as speed, power and number of processors used, datacenter's energy efficiency and energy mix~\cite{patterson2021carbon}. Studies have also focused on developing tools~\cite{lacoste2019quantifying, lannelongue2021green, schmidt2021codecarbon, faiz2023llmcarbon} to quantify energy consumption and carbon emission while running AI models using methodologically independent approaches that could be used by model developers, but have differing outputs that makes comparisons difficult. The lack of transparency by large AI model developers around energy consumption, their datacenter characteristics coupled with the lack of standards makes this task challenging. The International Energy Association (IEA) estimates energy demand from data centers and transmission to be about 1-1.5\% of global electricity use in 2023, but growing AI usage may increase pressure on energy resources~\cite{iea} which can undermine climate change mitigation goals. Consequently, the IEA identifies that ``more efforts are needed" in tracking usage from this sector, especially as power grids already face growing demand from population growth, and increased heating and cooling demands due to the worsening climate crisis.

AI also has enormous water consumption footprint with a high demand for clean freshwater to cool its datacenters. Water consumption from AI can be both on-site for the purpose of cooling datacenters as they run power hungry AI models, and also off-site during electricity generation~\cite{li2023making, zuccon2023beyond}, with current studies assessing water demand in information retrieval~\cite{zuccon2023beyond} and generative AI use~\cite{li2023making}. Analyses in this area is also restricted due to lack of transparency around water usage at various stage of AI development.
\begin{figure*}
\centering
\resizebox{5.5in}{3.2in}{\includegraphics{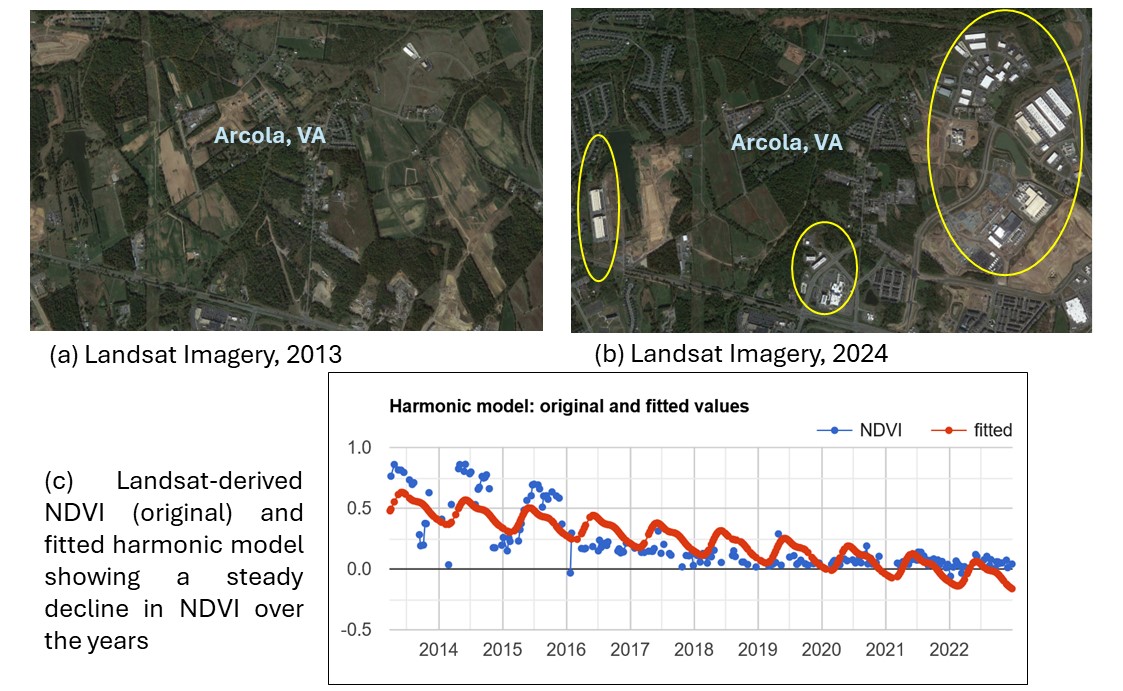}}
\vspace*{-.2mm}
\caption{Visible Landsat-8 imagery around Arcola in Northern Virginia showing changes in land use with comparison over a decade through (a) before and (b) after images. (c) Trend in the region's vegetation index (Normalized Vegetation Difference Index (NDVI)) shows a decline.}
\vspace*{-.5mm}
\label{fig:ndvi}
\end{figure*}

Moreover, the intersection of AI and environment is not limited to its operational energy and water use. Throughout AI's lifecycle, additional environmental impacts occur through embodied carbon emission -- that is associated with emission caused in manufacturing, fabricating, and recycling of computing equipment, which is under-reported, and embodied water footprint -- that is the water use associated with manufacturing and use over the lifecycle of the AI servers~\cite{gupta2021chasing, patterson2021carbon, li2023making}. Both embodied carbon emission and water use remain challenging to assess due to lack of transparency. Furthermore, there are impacts of AI in addition to this, such as land use change, alterations to air quality, heat dissipated to the surrounding environment and to facilities co-located with datacenters. Energy, carbon and the broader environment have been identified on the list of 100 factors around which more transparency is required from foundation model use~\cite{bommasani2023foundation}, yet AI risk assessment frameworks do not have clear guidelines and standards that should be required for assessing AI's environmental impact. While there is a need for transparent reporting on AI's resource consumption from technology companies, there is still a critical need for timely data on where, when, and what changes in the natural environment are occurring around the planet as a result of a surge in AI training and deployment. Such data will allow the impact assessment for accountability, extracting actionable insights informing sustainable planning and policy, and monitoring the impact of policies over time.  This study proposes a comprehensive monitoring methodology of AI's environmental footprint by (i) leveraging Earth Observation (EO) data (as they are global and openly available) to quantify changes occurring around datacenters over time, and (ii) building on existing recommendations and tools to create a framework as a step towards a holistic view of the intersection of AI and natural systems.

\section{Assessing the AI Environment Intersection}
The impact of AI on the physical environment is multifaceted, resulting in changes in emission, water availability, land use, air quality, and temperature. A large part of the AI-environment interactions is likely to occur at regions where AI has a physical footprint -- around its datacenters, and physical infrastructure. We propose a methodology to examine these locations using openly available, global, spatially explicit data for monitoring changes in key environmental variables over time. For our assessment we explore the potential carbon footprint and energy source around these physical sites wherever available. As these interactions are likely to play out over time across the different datacenter sites around the planet, there is a knowledge gap in what (type of change), when (time of change), and where (location) these changes are occurring -- one where Earth observation (EO) satellite data can fill in the existing gaps. We describe our spatially explicit analysis methodology below:

\subsection{Assessing Carbon Intensity and Clean Energy Around Datacenter Sites}
\label{ref:elec} We first explore the datacenter and cloud locations of the three large publicly available providers, Google, Amazon, and Microsoft based on their locations and examine the likely electricity source and carbon intensity of the major energy providers in that region. To estimate the electricity source and carbon intensity, we use yearly data from ElectricityMaps (\url{https://www.electricitymaps.com/}). This tracks data on electricity that flows on the power grid and is available across several countries based on zones categorized by electricity providers. In the United States, Google lists of total of 14 datacenters, Amazon with four zones (except government cloud), and Microsoft with eight locations. The average carbon intensity of power grid in the areas where these datacenters are located is approximately 411.71, 323.75, and 406.37 gCO$_{2}/$kWh respectively. Data from ElectricityMaps is unavailable over certain countries and can be extended elsewhere over the other sites. While this does not directly estimate the electricity consumption or emission caused by AI infrastructure, it represents the carbon intensity of the most widely available electricity provider in the region. As transmitting electricity over long distances is inefficient, and renewable energy (if purchased) may be only available sporadically, and with limitations in energy storage, AI datacenters are likely to use the most widely available energy source (a portion of which may be renewable) and this tool can be used to estimate the relative indicator of clean energy access across different sites.

\begin{figure*}
\centering
\resizebox{7in}{2.2in}{\includegraphics{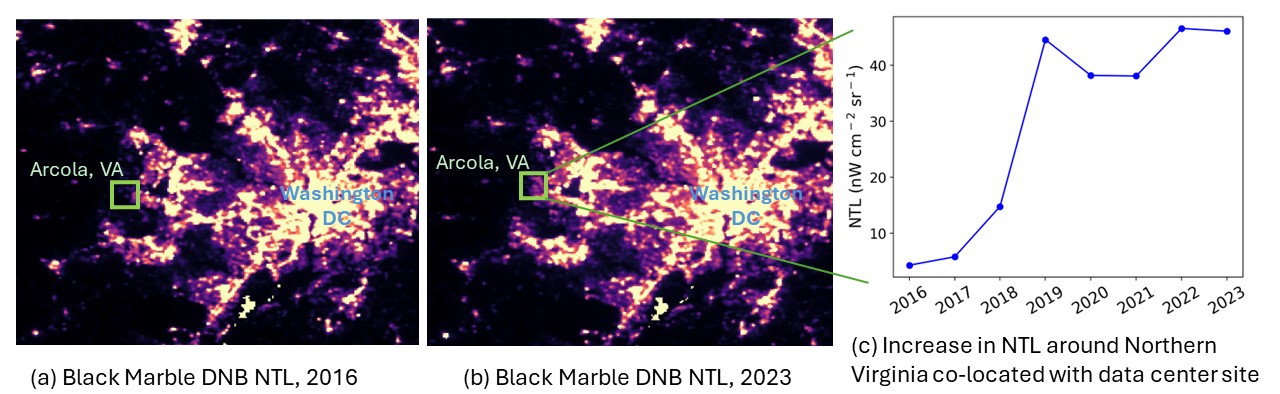}}
\vspace*{-.2mm}
\caption{Nighttime light (NTL) from VIIRS DNB shows urban infrastructure development and electricity usage. A comparison of DNB-derived NASA Black Marble annual composite data shows the increase in nightlight in the highlighted area around Arcola, VA over a span of eight years comparing (a) before and (b) after images. (c) A time-series of the annual NTL over the same area shows a very high (10X) increase in NTL levels.}
\vspace*{-.5mm}
\label{fig:ntl}
\end{figure*}

Data Gap and changes required: Data from ElectricityMaps is a available for several countries across the global that is updated at varying temporal scales and to the best of our knowledge represents the only openly available dataset describing regional clean energy characteristics of the grid that is available to datacenters based on their locations. However greater transparency from technology companies would be necessary around their datacenter and cloud infrastructure, their electricity providers, the energy mix, carbon intensity of backup generators, efficiency of datacenters reported through Power Usage Effectiveness (PUE), and access and usage of renewable energy. Microsoft Azure aggregated report from 2022 shows a global PUE of 1.18 (Americas: 1.17, Asia Pacific: 1.405, Europe, Middle East, Africa: 1.185), while individual datacenter PUE are reported for selected locations. Google currently reports a PUE of 1.10 in 2023 and lists site-specific breakdown. Datacenter-specific metrics by technology companies would provide greater transparency. The embodied emission of datacenters should also be reported. Similar metric on datacenters' water usage (Water Usage Effectiveness (WUE))are underreported (with the exception of selected Microsoft datacenters) and should be required for tracking water usage~\cite{li2023making}.
\subsection{Integrated Monitoring of Environmental Changes Around AI Datacenter Sites: A Case Study Around Northern Virginia}

Alterations to the natural environment due to increased AI infrastructure presence and operations can impact land (land use change, local ecosystem), atmosphere (carbon fluxes as a result of land cover transformation, dissipated heat, emission at power generation sites and changes to local air quality), increased water consumption, to name a few. Along with tracking datacenter-specific metrics and efficiency reports, monitoring local impact of changes at the AI infrastructure sites are necessary which requires globally available, spatially explicit data over time. Satellite observations provide planetary-scale data over time and monitor key environmental variables and is available openly. As a case study, we explore the role satellite data can play in observing changes over time and select the region around Northern Virgina, United States as it hosts the largest market for datacenters.

As the explicit locations of the datacenters are unknown, we use the existing and proposed data center map curated by the Piedmont Environmental Council (PEC)~\cite{n-va} that is crowd-sourced and being updated. We note the sites of existing datacenters around Arcola, VA, a region that is facing a growth in datacenter development and lies around eight miles south of Ashburn, VA, also termed as data center alley. The PEC map marks a total of 12 datacenters in this region, with six existing and six proposed sites. Out of the six existing sites, the map notes three Amazon datacenters, and a Mircosoft datacenter.
We first explore visible imagery from a Landsat-8 composite for observing any visible changes and any changes in nighttime lights from NASA Black Marble annual composite around Arcola. The visible imagery comparison is shown in Figure~\ref{fig:ndvi}(a) and (b) and marks the development that has occurred in the region. Figure~\ref{fig:ntl}(a) and (b) shows the changes in nightlights and shows the regions' proximity to other major metropolitan areas. Nightlights have been an indicator for urban development and infrastructure, power usage and are even used to estimate anthropogenic carbon emissions. We observe an increase in nightlight intensity in the Arcola region (marked by the green box).
We then analyze some key environmental variables around the known datacenter sites and observe the following environmental indicators and their changes:
\begin{itemize}
    \item Normalized Difference Vegetation Index (NDVI): This is a commonly used vegetation index, derived here from Landsat-8 composite using the reflectance in near infrared and red wavebands for a decade at 30 m spatial resolution. NDVI relates to the greenness of the region (higher is better). We used Landsat-8 data from U.S. Geological Survey (USGS)~\cite{landsat} using USGS EarthExplorer and Google Earth engine and used quality layers to select high quality, cloud free observations that results in a sparse time-series as seen in Figure~\ref{fig:ndvi}(c) (marked in blue). As vegetation shows strong seasonal component, harmonic model fitting is commonly used to estimate the NDVI time-series and characterizes regional properties~\cite{jakubauskas2001harmonic, lhermitte2008hierarchical, chakraborty2018time, zhu2020continuous}.  We apply this approach to model NDVI time-series by decomposing it as a series of sinusoids, each with its unique amplitude and phase using
\begin{equation}
    \hat{\rho}_t=\mu + \alpha_{1}cos\Big(\frac{2 \pi}{N} t \Big)+ + \alpha_{2}sin\Big(\frac{2 \pi}{N} t \Big) +\epsilon,
    \label{eqn:refmodel}
\end{equation}

where $\mu$ represents the mean NDVI component, $\alpha_{1}$ and $\alpha_{2}$ are the seasonal components of NDVI, N=365 shows the days in a year, $\epsilon$ represents the error term and $\hat{\rho}_t$ is the model predicted NDVI. The parameters for NDVI time-series is estimated from available Landsat-8 observations using ordinary least squares (OLS) regression. Using the estimated NDVI $\hat{\rho}_t$, we observe a steady decline in the vegetation index over the decade in the region in Figure~\ref{fig:ndvi}(c) (marked in red). Google cloud operations in this region are noted to have begun in 2018, with the datacenter operations from 2019. Amazon launched operations in this zone starting 2006, while Microsoft in 2012-2014. The observed decline in NDVI cannot be attributed to AI and datacenters alone, but the index offers a way to monitor the natural environment at the sites of anthropogenic and AI activity hotspots. These changes in vegetation can impact atmospheric fluxes, evapotranspiration, temperature and alter local environmental properties. Satellite data offers a way to monitor vegetation condition around AI datacenter sites and map ongoing land cover alterations in this framework.

\item Nighttime Lights (NTL): NTL derived from remote sensing data from the Day/Night Band (DNB) onboard the Visible Infrared Imaging Radiometer Suite (VIIRS) on the Suomi-NPP satellite capture urban infrastructural dynamics relating to human and environmental changes~\cite{roman2018nasa, levin2020remote}. NTL observations have been used to monitor infrastructural changes due to urbanization~\cite{xie2019temporal, chakraborty2023adaptive}, disasters impacting power grids~\cite{roman2019satellite,chakraborty2023adaptive}, and are used to estimate anthropogenic emission~\cite{oda2018open}. The DNB captures visible nighttime lights from natural as well as human-generated sources such as city lights providing a global view of the power grid from space~\cite{roman2018nasa}. NASA's Black Marble annual product (VNP46A4) tracks the annual changes in NTL derived from the DNB at a 500 m spatial resolution and we downloaded the image tile corresponding to this region from NASA’s Level-1 and Atmosphere Archive and Distribution System Web Interface (LAADS-DAAC). We extracted the NearNadir-Composite-Snow-Free layer from the annual composite.
We observe the annual decade-long changes in the region in NTL as seen in Figure~\ref{fig:ntl} (c). A 10X increased in NTL around datacenter sites in Arcola is observed, indicative of increased economic activity, urbanization and urban infrastructure use, with an exception in 2020-2021 that could be related to the economic activity decline due to COVID-19. The increased electricity consumption results in emission (regional power grid with carbon intensity of 430gCO$_{2}/$kWh and 39\% low-carbon only 7\% renewable energy based on ElectricityMaps data) and light pollution. Similar to vegetation, not all nightlight changes can be attributed to datacenters, but serves as an important metric in monitoring the natural environment around AI compute resources. Moreover, NTL derived datasets such as the Open-Data Inventory for Anthropogenic Carbon dioxide (ODIAC) may also be consulted, albeit at the monthly temporal resolution, to estimate carbon emission around these sites.
\begin{figure}
\centering
\resizebox{3.5in}{1.45in}{\includegraphics{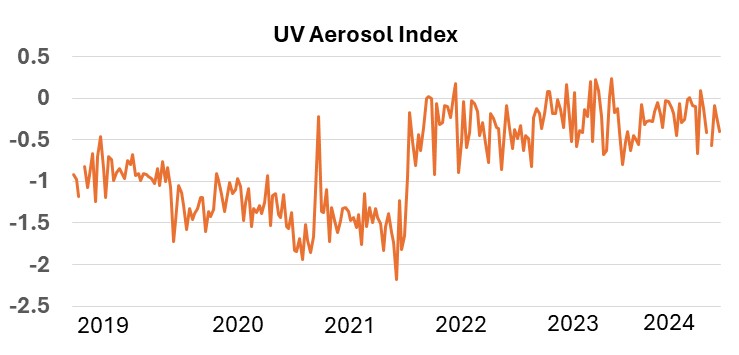}}
\vspace*{-.2mm}
\caption{TROPOMI derived measure of ultraviolet aerosol index showing aerosol properties over the Arcola, VA region through time. Atmospheric composition data will be an important measure to track change in air quality due to land cover changes and emission from increased energy consumption.}
\vspace*{-.5mm}
\label{fig:aerosol}
\end{figure}
\item Atmospheric Composition: Increased anthropogenic activity expected around AI datacenter sites are likely to introduce variations in atmospheric composition. The TROPOspheric Monitoring Instrument (TROPOMI) satellite on board the Copernicus Sentinel-5 Precursor satellite measures various atmospheric species related to air quality and climate since 2018 at a spatial resolution of 3.5 km. We use the TROPOMI time-series measurements of ultraviolet aerosol index (UVAI)~\cite{tropomi-uvai} for exploratory data analysis around the datacenter sites using the TROPOMI explorer~\cite{tropomi}. The UVAI shows a gradual increase as shown in Figure~\ref{fig:aerosol}. UVAI shows presence of aerosol, dust, clouds that impact Earth's energy budget~\cite{stier2006impact, yu2006review}. UVAI variations indicates changes in aerosol and atmospheric composition ~\cite{tropomi-uvai} and can be an important metric to interpret regional air quality changes. 
Due to the dynamic nature of atmospheric processes, complementing this analysis with atmospheric transport models and source attribution may be necessary to determine the contributors to this increase. This assessment is also necessary at the power grids supplying AI datacenters.
\end{itemize}
\textbf{Future extension}: The metrics analyzed above show the usefulness of satellite observations in capturing trends in environmental variables around an AI activity hotspot and when extended, can serve as critical global measures tracking changes occurring around datacenters. Particularly, given the variety of datasets, these datasets can show what changes are occurring -- for example decline in vegetation, increase electricity consumption and light pollution from NTL, variation in atmospheric properties, and when and where they occur. As this is an exploratory case study, we used the datasets described above to demonstrate the usefulness of satellite based measures. In future extensions, we will extend this to a wider range of datasets such as atmospheric composition (TROPOMI NO$_{2}$, ozone, methane), carbon emission (from Orbiting Carbon Observatory), temperature, additional vegetation metrics (such as Enhanced Vegetation Index), satellite maps of oil and gas infrastructure~\cite{omara2023developing}, water quality data, biodiversity, etc., for a comprehensive insight into the environmental changes and key resources (energy, oil and gas maps, water) around these locations.
Moreover, the interconnectedness between land, water and energy is complex and the collective assessment using a range of datasets is critical to assess the technology-environment interface~\cite{cremades2019ten}. The datasets used in the case study and the ones identified for the expansion of this approach are globally available and is therefore scalable to the other sites. These datasets are acquired at varying resolution, and it may be necessary to map AI impact assessment needs to the appropriate environmental variable and resolution.

\textbf{Limitations, Planning and Policy Implications}: Earth observation data can thus provide insight into what changes are occurring over time around each datacenter site. We acknowledge that these changes cannot be attributed to AI datacenters alone, unless the sites are isolated from other anthropogenic activities. Moreover, datacenter use is not just limited to AI but hosts a collective range of technologies. However, these analyses can provide insights into transformations occurring around sites with datacenter expansion and inform sustainable planning policies. Additionally, raw material extraction for AI processors that may cause environmental degradation can also be monitored using EO data. Furthermore, transparency from technology companies around AI model training, deployment and inference (time, processor specifications, power and water usage) can provide an estimate on the fraction of datacenter usage attributed to AI, and in turn approximate its contribution to these observed environmental changes.

Data Gaps and changes required: The primary data gap here is the precise knowledge of datacenter sites globally of different providers. This information can be shared as ``controlled release" with AI and environmental auditors and researchers, while the impact assessment can be shared publicly. Stakeholders such as local government, environmental councils, and residents who are directly impacted by these transformations should also be given access and agency over these datasets. Assessing environmental transformation requires earth science knowledge and expertise. Earth science researchers with expertise in these sub-fields should also be involved in assessing satellite-based insights and in understanding the impact of changes occurring around AI sites. 

\subsection{Gaps and Recommendation for AI-Environment Impact Assessment}
A key factor that has enabled AI's performance and pervasiveness is the compute resources derived from the natural environment. These resources are used throughout AI infrastructures' lifecycle and can adversely impact the environment. This interaction has remained under explored and requires attention to stay on track with the climate change mitigation effort towards the 1.5 °C goal under the Paris Climate Agreement. Given the multifacted interaction and complex feedback between various Earth system components, this study identifies the following key information gaps and steps to address them to assess the impact of AI:
\begin{enumerate}
    \item Technology companies should be required to be transparent around electricity and water sources and consumption, datacenter-specific efficiency metrics such as PUE, WUE, energy mix and carbon intensity of electricity used, access and use of renewable energy, and energy, water consumption and emission throughout the lifecycle of AI processors and datacenters. These will increase trust in companies' sustainability commitment, allow periodic assessment of consumption and identify best practices for optimal resource usage which can also reduce the cost incurred by AI companies to acquire and maintain these resources.
    \item Technology companies should be required to be transparent around processors used, and their speed and power. These statistics can be incorporated into ML carbon emission tools for use by AI model developers at various stages of the AI pipeline (training, finetuning, inference) and can reduce the differences in carbon footprint estimated by the tools~\cite{bannour2021evaluating}.
    \item Alignment of AI evaluation with environmental footprint measures should be given priority. With the increasingly competitive AI market, attention is often given to common model performance metrics on different ML tasks. Evaluating models based on energy usage and requiring fewer FLOPS, if embraced by the ML community, could be a step towards more energy efficient models~\cite{miao2023towards, zheng2024learn}. Additionally, technological advancements in energy efficiency and renewable energy storage, transmission can also be crucial in lowering AI's footprint.
    \item Comprehensive insight into what transformations are occurring around AI infrastructure sites (including datacenters, their electricity and water sources, backup generators) should be derived from global, openly available satellite datasets as demonstrated in this study. Assessment of these changes over time and its impact on environment, human health (excess heat, poor air quality), and biodiversity should be studied in collaboration with earth scientists. These effects should be used to inform policy frameworks and planning. This could result in an AI Environmental Impact inventory similar to emission inventories and compiled by companies across its operation sites. EO datasets can also play a role in future datacenter site selection and examining various tradeoffs, for example between availability of solar energy and water demand for cooling~\cite{li2023making}.
    \item Greater transparency and public education around AI's environmental cost is needed. At present, the energy and water cost of various AI models remain challenging to estimate while the interest in easily accessible AI tools remain high among millions of end users. Understanding the hidden cost of running these tools may create awareness around when using this technology is justified. For example, the IEA estimates that datacenters' energy consumption will double from 2022 to 2026, equalling Japan's energy consumption~\cite{iea-proj}, while approximate 660 million people still lack access to electricity based on UN reports~\cite{un-elec}.
\end{enumerate}
The proposed steps require coordinated effort and regular audits to be successful and should be updated over time as the technology landscape evolve.

\section{Conclusion}
Rapid progress in artificial intelligence (AI) offers a wide range of opportunities. However, currently there is limited understanding on how, where and when this technology is impacting the natural environment, which supports its compute and resources needs. In this study we assess the various environmental areas impacted by AI. We identify the need for transparency and spatially disaggregated analysis around datacenter locations. For this, we propose the use of spatially explicit global data that is openly available for scaling the analysis across different sites such as those acquired by satellites. We demonstrate the usefulness of satellite data and our methodology in tracking environmental changes using a case study around known datacenter locations in Northern Virginia, United States. We observe noticeable changes in environmental metrics around the sites and identify the next steps in expanding this methodology for a more holistic assessment. We also identify some key data gaps and steps needed collectively by AI companies and researchers to comprehensively monitor AI's environmental impact .

\bibliographystyle{unsrt}  
\bibliography{references}

\end{document}